\documentclass[aps,pra,floatfix,amsmath,showpacs,lengthcheck, superscriptaddress]{revtex4-1}
\usepackage{graphicx}
\usepackage[utf8]{inputenc}
\usepackage{amssymb}
\usepackage{url}
\usepackage[colorlinks=true,%
urlcolor=blue,% 
linkcolor=blue,% 
citecolor=blue,% 
hypertexnames=true,%
]{hyperref}% 
\begin{document}

\title{Simulation of Hot-Carrier Dynamics and Terahertz Emission in Laser-Excited Metallic Bilayers}

%%%%%%%%%%
\author{Dennis M. Nenno}
%\email{nenno@physik.uni-kl.de}
\affiliation{Physics Department and Research Center OPTIMAS, University of Kaiserslautern, 67663 Kaiserslautern, Germany}
\affiliation{College of Optical Sciences, University of Arizona, Tucson, AZ 85721, USA}
\author{Rolf Binder}
\affiliation{College of Optical Sciences, University of Arizona, Tucson, AZ 85721, USA}
\author{Hans Christian Schneider}
\email{hcsch@physik.uni-kl.de}
\affiliation{Physics Department and Research Center OPTIMAS, University of Kaiserslautern, 67663 Kaiserslautern, Germany}


\date{\today}

\begin{abstract}
Metallic bilayer structures have been shown to emit strong terahertz (THz) pulses. We present a predictive multiscale mode that simulates optically induced spin-currents in a Fe/Pt-heterostructure and the emitted electric field. Electronic effects are treated on the nano scale using the Boltzmann transport equation for the dynamics of out-of-equilibrium charge carriers, numerically solved with a particle-in-cell code. The optical effects are simulated with a formalism that bridges the nanometer scale of the structure to the micrometer scale of the emitted waves. The approach helps to understand recent experimental findings on the basis of microscopic scattering effects and transport phenomena. Our theory's versatility allows it to be readily adapted to a wide spectrum of spintronic THz emitter designs. As an example, we show how the THz generation efficiency, defined as output to input power ratio, can be improved and optimized using serially stacked layers in conjunction with THz anti-reflection coatings.
\end{abstract}

\maketitle

%%%%%%%%%%%%%%%%
% Introduction %
%%%%%%%%%%%%%%%%
\section{Introduction}

Nanoscale spintronic structures excited by ultrashort optical pulses provide a novel emitter technology for broadband terahertz (THz) radiation~\cite{Kampfrath:2012kw,Seifert:2016kc}. Emission strength and efficiency has been intensively studied in structures of various materials and the generated pulses exceed the power of conventional THz emitters. As they are comparatively cheap to fabricate~\cite{Seifert:2016kc}, they open up new possibilities to study intricate material parameters~\cite{acs.nanolett.7b04538}. In metallic heterostructures, the emission can be traced back to a charge current perpendicular to the layer stacking in the non-magnetic part of the slab~\cite{Seifert:2016kc}. This charge current is generated by an optically induced spin-current through the inverse spin-Hall effect. The structure then acts as an antenna for these oscillating charge-currents and emits a short pulse with broad spectrum in the THz range. Subsequent studies found a strong dependence of the emitted pulse strength on the spin-Hall angle of the non-magnetic material~\cite{Seifert:2016kc}. Both the thickness dependence and the laser excitation wavelength in Fe/Pt structures have been studied as well~\cite{Torosyan:2018iv,2018arXiv180308838P,2018arXiv180800746H}.

Currently, there is no complete model to describe all the underlying processes from excitation to emission. Experimental studies concentrate on extracting the correct shape of the spin-current from the measured field pulse to compare with simulations~\cite{Seifert:2016kc}. Other studies have used simplified approaches to extract the spin-diffusion length that describes the decay of the optically induced spin-currents~\cite{Torosyan:2018iv}.

The purpose of this paper is to present a model that simulates the carrier dynamics after ultrashort laser excitation using the Boltzmann transport equation (BTE) in Fe/Pt multilayers~\cite{Nenno:2016cm}. The BTE has been proven to be an adequate tool to simulate excited carrier dynamics in metallic structures on the nanoscale~\cite{Hurst:2018bz,Manfredi:2005ba,Nenno:2016cm,Nenno:KNmjgCOZ}. To solve for the carrier-mediated spin-current, we use the particle-in-cell approach that introduces numerical superparticles to represent the hot-carrier distribution. Calculations of field propagation are both used as an input to estimate the excitation strength as well as to calculate spectrum and shape of the emitted pulse. Thus, we account for both laser absorption within the layers and the subsequent dynamics as well as for the propagation of the emitted fields through the structure.

%%%%%%%%%%%%
% Figure 0 %
%%%%%%%%%%%%
\begin{figure}[t]
	\centering
	\includegraphics[width=0.45\textwidth]{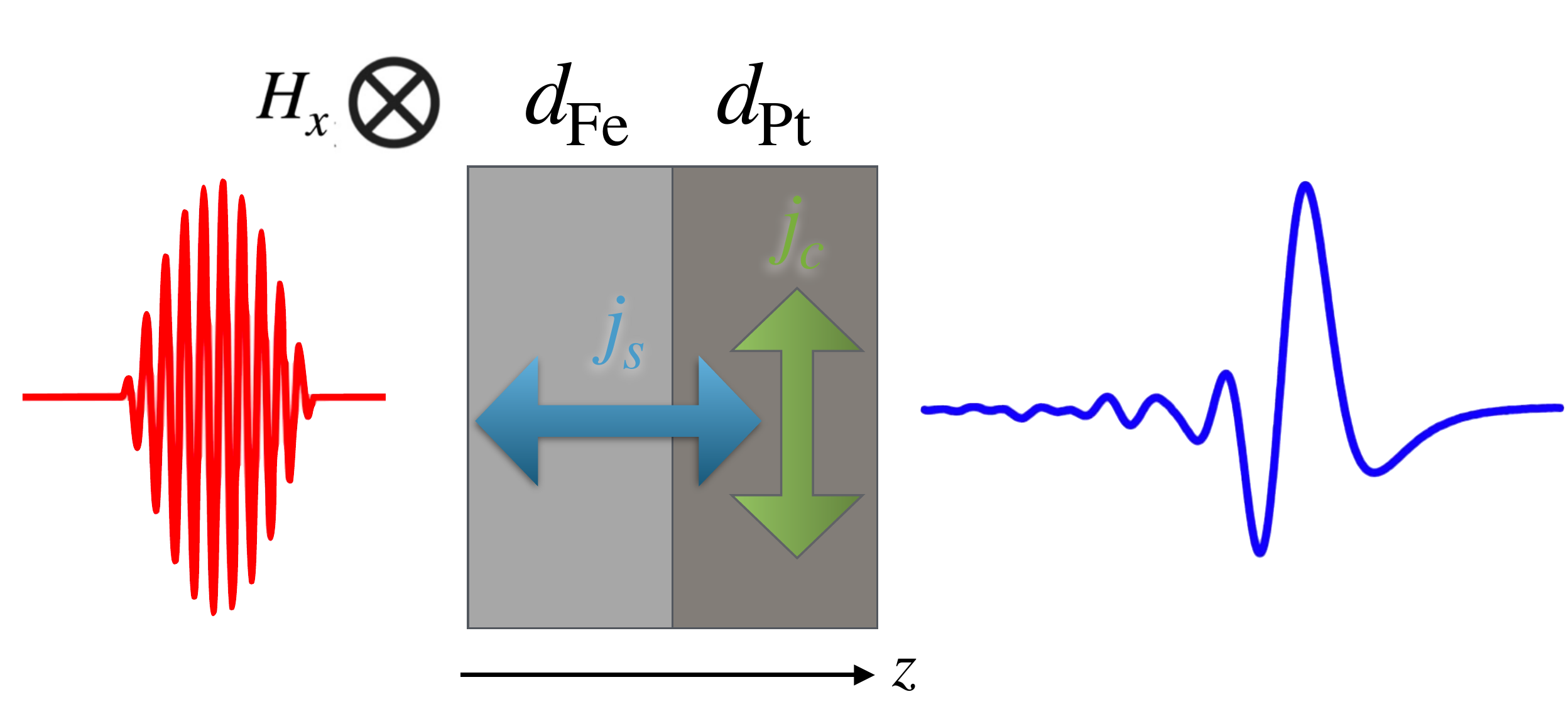}
	\caption{Typical set-up of a spintronic terahertz emitter structure with indicated laser excitation and emitted pulse. The laser-induced spin current $j_s$ as well as the resulting charge current $j_c$ due to the inverse spin-Hall effect are shown.\label{Fig0}}
\end{figure}

%%%%%%%%%%%%%
% Figure 1a %
%%%%%%%%%%%%%
\begin{figure*}[t]
	\centering
	\includegraphics[width=\textwidth]{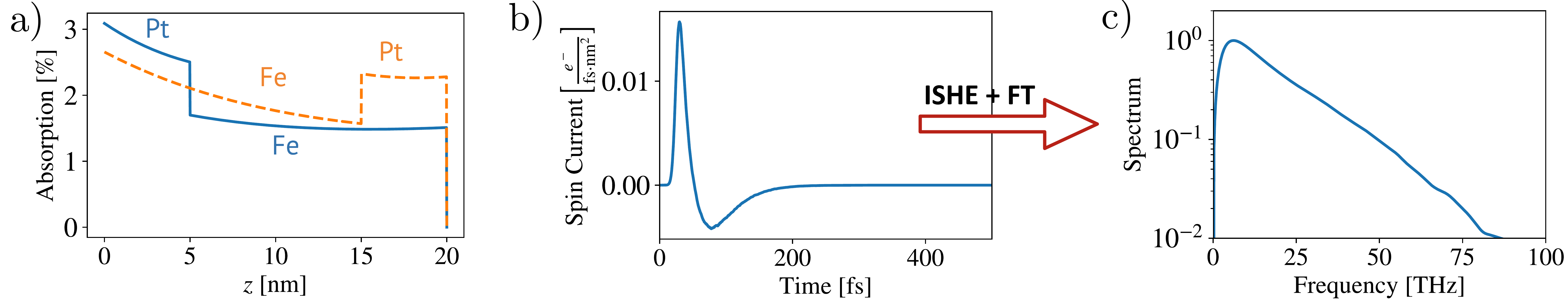}
	\caption{The absorption as a fraction of incident power for 800\,nm s-polarized light in a Fe(15\,nm)Pt(5\,nm) layer is plotted (a) for irradiation from the Pt-side (solid) and from the Fe-side (solid) and normal incidence. The optically induced spin current (b) and the spectrum of the charge current (c) are shown after excitation with a 10\,fs infrared laser pulse. \label{Fig1a}}
\end{figure*}

The paper is organized as follows. We describe our approach to the laser absorption in the individual metal layers. We then introduce the guiding equation for the carrier dynamics, sketch its numerical solution and introduce the inhomogeneous Helmholtz equation for the emitted optical fields. We then present results for the induced spin-currents in a Fe/Pt heterostructure and the structure of the emitted field on the example of varying laser pulse duration. We show that the emitted field shape is directly proportional to the shape of the current. In the last part of the paper, we propose a serial stacking of spintronic emitters involving anti-reflective coatings to increase mono-energetic output in a feasible way.

\section{Model}
%%%%%%%%%%%%%%%%%%%%%%%%%%%%%%%
% Theory - Optical Absorption %
%%%%%%%%%%%%%%%%%%%%%%%%%%%%%%%
We organize the different theoretical approaches presented in this section by the temporal order of the physical processes involved in the terahertz generation. All three parts, excitation, electron dynamics and emission as well as the structure are sketched in Fig.~\ref{Fig0}. First, an optical laser pulse irradiates the structure.

Hence, in a first step we calculate the number of excited carriers over the depth of the metallic bilayer. These excited carriers then traverse the structure, scatter and induce a spin current into the non-magnetic metal. The dynamics are described in a second step using a Boltzmann transport approach. Due to the inverse spin-Hall angle, the spin current induced in the platinum layer is converted into a perpendicular charge current. In the final step, we will use the calculated charge current as an input to solve the inhomogeneous Helmholtz equation, which we solve on the millimeter scale to obtain the emitted field and its spectrum.

\subsection{Laser Absorption}
The absorbed laser power is calculated for an excitation at 800\,nm wavelength, which corresponds to laser wavelengths typically used in experiments~\cite{2018arXiv180800746H,2018arXiv180308838P,Torosyan:2018iv}.

We use a standard finite-difference approach to solve the Helmholtz equation for the complex field amplitudes $\mathcal{E}$ and $\mathcal{H}$, denoting electric and magnetic field respectively. The boundary condition is an incoming wave at $\lambda_0 = $ 800\,nm. From the resulting fields, we calculate the negative derivative of the absorbed power in one layer $m$~\cite{Prentince_JOPD_2000},
\begin{equation}
A_m(z) = -\frac{dP_m(z)}{dz} \, ,
\label{eq:absorbtion}
\end{equation}
where the amplitude of the Poynting vector is given by
\begin{equation}
P_m(z)=\mathrm{Re}[\mathcal{E}_m(z)\mathcal{H}^*_m(z)]
\end{equation} 
and is normalized to the amplitude of the incident field. The field amplitudes for $\mathcal{E}$ and $\mathcal{H}$ are given by $\mathcal{E}^+_m(z)+\mathcal{E}^-_m(z)$ and $\tilde{n}_m(\mathcal{E}^+_m(z)-\mathcal{E}^-_m(z))$ respectively. The right- ($\mathcal{E}^+$) and left-propagating ($\mathcal{E}^-_m$) contributions to the field can be easily extracted from the calculation by comparing field value and spatial derivative. The complex index of refraction $\tilde{n}_m$ at the laser wavelength of 800\,nm is taken from Ref.~\onlinecite{doi:10.1063/1.3243762} for iron and platinum. The total absorbance of the structure is given by the sum over the contributions from individual layers~\cite{Prentince_JOPD_2000},
\begin{equation}
	A = \sum_m \int_{0}^{d_m} A_m(z)\,dz \, ,
\end{equation}
where the integral runs over the thickness $d_m$ of each layer.
For structures that are irradiated through a material with a refractive index that lies between that of Fe/Pt and vacuum, e.g.,~the frequently used substrate MgO, the obtained values for the total absorbed energy can be higher~\cite{Torosyan:2018iv}.
 
Integrating over the total length of a Fe(10\,nm)Pt(5\,nm) slab, 41\% of the laser pulse energy is absorbed when irradiated from the iron interface and 37\% from the platinum interface. The depth-dependent absorption relative to the incident laser power is shown in Fig.~\ref{Fig1a}(a).

Through the absorbed energy over the duration of the pulse, we can estimate the number of excited electrons as a function of position, assuming  a linear relation for the generation rate~\cite{PhysRevLett.85.844}.
The position-dependent hot-carrier number is then used as an input for the simulation of the electron dynamics.

%%%%%%%%%%%%%%%%%%%%%%%%%%%%%%
% Theory - Electron Dynamics %
%%%%%%%%%%%%%%%%%%%%%%%%%%%%%%

\subsection{Electron Dynamics}
%We use the calculated number of excited carriers as an input for the following calculation of the hot-carrier dynamics. 
While the electromagnetic field propagates through the structure, electrons are excited to unoccupied states above the Fermi level. These hot carriers then move through the structure and relax back to equilibrium. This electronic dynamics is often called superdiffusive transport and we describe it with a Boltzmann transport equation~\cite{Battiato:2010br,Nenno:2016cm}. On the femto- to picosecond timescale, we consider two types of scattering processes. Scattering events with impurities and phonons are considered elastic. Secondary-carrier generation due to hot-electron scattering with equilibrium carriers is included using inelastic carrier scattering times~\cite{Nenno:2016cm}.

\begin{figure}[b]
	\centering
	\includegraphics[width=0.35\textwidth]{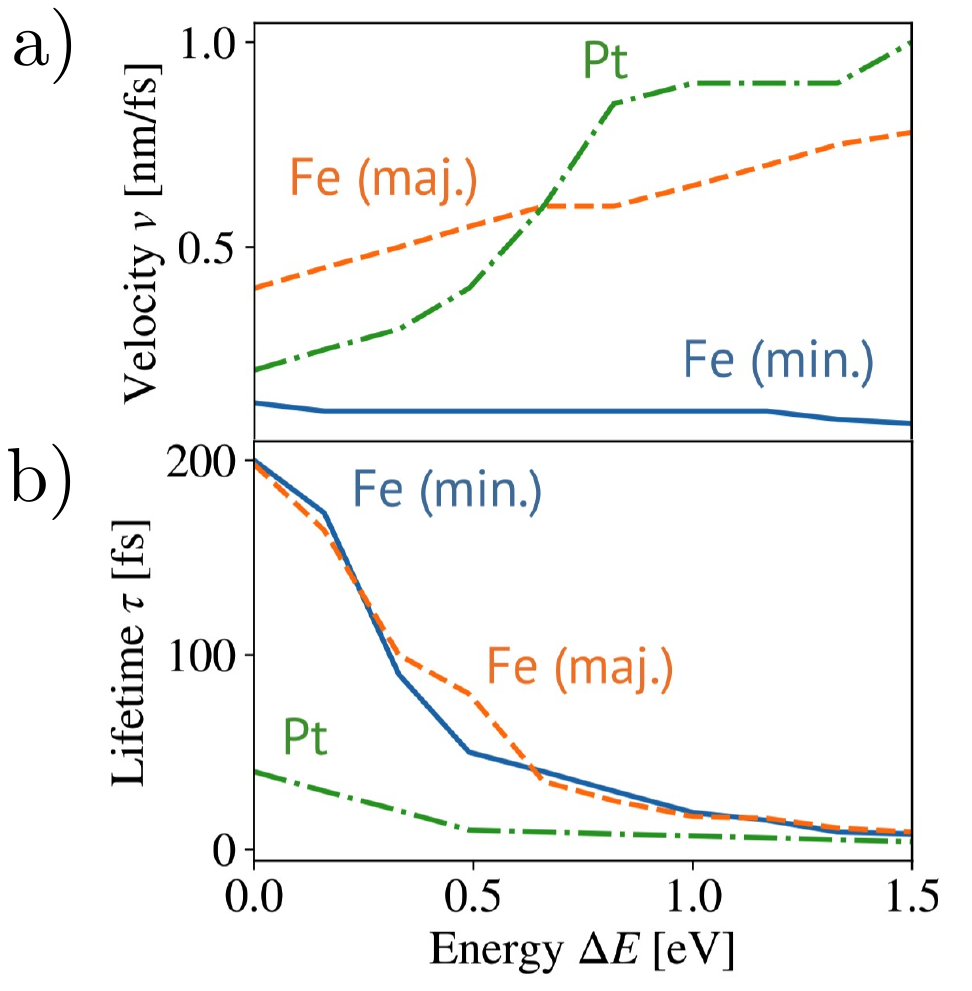}
	\caption{The electron dynamics is described using \textit{ab initio} input for carrier velocities (a) and lifetimes (b) above the Fermi level. Data is taken from Ref.~\onlinecite{Zhukov:2006ky} for $v_{\mathrm{Pt}},v_{\mathrm{Pt}}$ and $\tau_{\mathrm{Pt}}$ as well as Ref.~\onlinecite{Kaltenborn:2014du} for $\tau_{\mathrm{Fe}}$.\label{Fig1b}}
\end{figure}

The formal treatment of electronic dynamics in the framework of the BTE starts with the distribution of hot carriers, $g_{\sigma}(z,E,\theta,t)$, which depends on spin $\sigma$, position $z$ in the slab, the particle energy $E$, its propagation angle with respect to the $z$-axis, $\theta$,  and time $t$. We assume a universal polarization axis along which the spin aligns parallel or antiparallel to, such that $\sigma = \uparrow, \downarrow$. The evolution of this distribution function is given by~\cite{Nenno:KNmjgCOZ}
\begin{widetext}
\begin{equation}
\begin{split}
\bigg[\frac{\partial}{\partial t} + v_\sigma(E)\cos(\theta) \frac{\partial}{\partial z}\bigg]& g_{\sigma}(z,E,\theta,t) = S_{\sigma}(z,E,t) - \frac{g_{\sigma}(z,E,\theta,t)}{\tau^{\mathrm{eff}}_{\sigma}(z,E)} \\
+
&\sum_{\sigma'}\int\frac{d\Omega'}{4\pi}\int dE'\,w(z,\sigma',E';\sigma,E)g_{\sigma'}(z,E',\theta',t)\rho_{\sigma'}(z,E') \, ,
\label{eq:bte2}
\end{split}
\end{equation}
\end{widetext}

where we denote the carrier velocity by $v$ and the source term for excitation processes by $S$. The last two terms in Eq.~\eqref{eq:bte2} describe out- and in-scattering processes due to interactions with equilibrium carriers and many-particle excitations. The effective, spin-dependent lifetime is given by
\begin{equation} \tau^{\mathrm{eff}}_{\sigma}(E)=1/(\tau^{-1}_{\mathrm{el}}+\tau^{-1}_{\sigma}(E))\, ,
\end{equation}
combining the elastic lifetime $\tau_{\mathrm{el}}$ and the inelastic scattering lifetime $\tau_{\sigma}(E)$. The integration over spin $\sigma$, energy $E$ and solid angle $\Omega$ in Eq.~\eqref{eq:bte2} distributes electrons in momentum space and includes secondary carrier generation. All scattering events are assumed to be local. We use material parameters from \textit{ab initio} calculations presented in ~\onlinecite{Zhukov:2006ky} and Ref.~\onlinecite{Kaltenborn:2014du} and shown in Fig.~\ref{Fig1b}. The numerical scheme to solve Eq.~\eqref{eq:bte2} is presented in Ref.~\onlinecite{Nenno:KNmjgCOZ}.
From the hot-electron distribution, we calculate the hot-carrier spin-current density
\begin{align}
	\begin{split}
		j_s(z,t) = \int \frac{d\Omega}{4\pi}&\int dE\,v(E)\cos(\theta)\\ \times& \left[g_{\uparrow}(z,E,\theta,t)-g_{\downarrow}(z,E,\theta,t)\right]\, .
	\end{split}
\end{align}
The charge current density resulting from the inverse spin-Hall effect is calculated by
\begin{align}
\mathbf{j}_c = \gamma \mathbf{j}_s \times \frac{\mathbf{M}}{|\mathbf{M}|} \, ,
\label{eq:spinhall}
\end{align}
following Ref.~\onlinecite{Seifert:2016kc}. Here, the spin-Hall angle is denoted by $\gamma$. In the following we assume a geometry shown in Fig.~\ref{Fig0}. The magnetization points in positive $x$-direction, so that Eq.~\eqref{eq:spinhall} is simplified to $j_c = \gamma j_s$, with $j_c$ denoting the magnitude of the charge current in $y$-direction. The spin current $j_s$ flows in $z$-direction. The current that is driven by the charge to spin conversion is then used to calculate the emitted field.

%%%%%%%%%%%%%
% Figure 1b %
%%%%%%%%%%%%%
%%%%%%%%%%%%%%%%%%%%%%%%%%%%%
% Theory - Optical Emission %
%%%%%%%%%%%%%%%%%%%%%%%%%%%%%

%%%%%%%%%%%%
% Figure 2 %
%%%%%%%%%%%%
\begin{figure*}
	\centering
	\includegraphics[width=502pt,height=116pt]{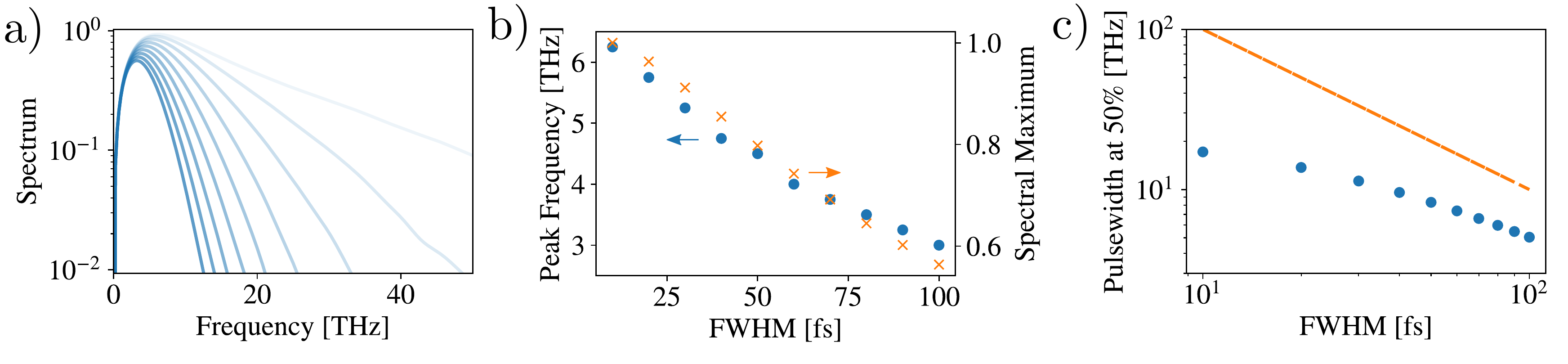}
	\caption{Emitted spectrum for different duration of the exciting laser pulse (10\,fs (dark blue) -- 100\,fs (light blue) in steps of 10\,fs) at 800\,nm wavelength (a). Position of the spectral maximum (dots) and its value (crosses) normalized to the strongest pulse (b). Spectral width at 50\% of the maximum value (dots) and comparison to the simple approximation $1/$FWHM (solid).\label{Fig2}}
\end{figure*}
\subsection{Terahertz Emission}
We determine the electric field $\mathcal{E}(z)$ emitted by the time-varying charge current in a metal layer using the inhomogeneous Helmholtz equation 
(see, e.g.,  p.~6 of Ref.~\onlinecite{meystre-sargent.91})
\begin{align}
\left[ \frac{d^2}{dz^2}+\frac{\omega^2}{c_0^2}\tilde{n}_{\mathrm{THz}}(z)^2\right] \mathcal{E}(z,\omega) =- i \mu_0 \omega j_c(z,\omega) \, ,
\label{eq:Helmholtz}
\end{align}
where $\omega$ denotes the THz photon frequency, $c_0$  the vacuum speed of light, $\mu_0$ the vacuum permeability,  $\tilde{n}_{\mathrm{THz}}(z)$ the complex refractive index of the materials at THz frequencies,
and $j_c$ is given by Eq.~\eqref{eq:spinhall}.
We stress that this is a completely different calculation from that for the absorption of the optical field, as typical thicknesses of spintronic emitters are on the order of nanometers, whereas the emitted THz waves have  wavelengths in the micro to sub-millimeter range. Thus, on the THz wavelength scale, the charge current can be treated in the  delta-layer approximation, $j_c(z,\omega) = \mathcal{J}_0(\omega) \delta(z-z_0)$.
The emitted field adjacent to the delta-layer then obeys the condition
\begin{align}
\mathcal{E}(z_{-}, \omega) =  \mathcal{E}(z_{+},\omega) = -\frac{1}{2} \mu_0 c_0 \mathcal{J}_0(\omega) \, ,
\end{align}
where $z_{\pm}= z_0 \pm \eta$  ($\eta \downarrow 0$) is on the vacuum side adjacent to each vacuum-metal interface.
This shows that the spectrum $I(\omega) \propto |\mathcal{E}(\omega)|$ and the temporal profile of the emitted field are directly proportional to the spectrum and profile of the current density, respectively.
We note that the delta-layer approximation is a convenient tool to bridge nanoscale to millimeter-scale problems, allowing one to simulate more complex structures as used in experiments, in which dispersive elements (e.g.\ MgO substrate and detector components) modify the spectrum. In this approach, conventional optical transfer matrix techniques can easily be combined with the transfer matrix for the delta-layer source.

%%%%%%%%%%%%
% Figure 3 %
%%%%%%%%%%%%
\begin{figure*}
	\centering
	\includegraphics[width=\textwidth]{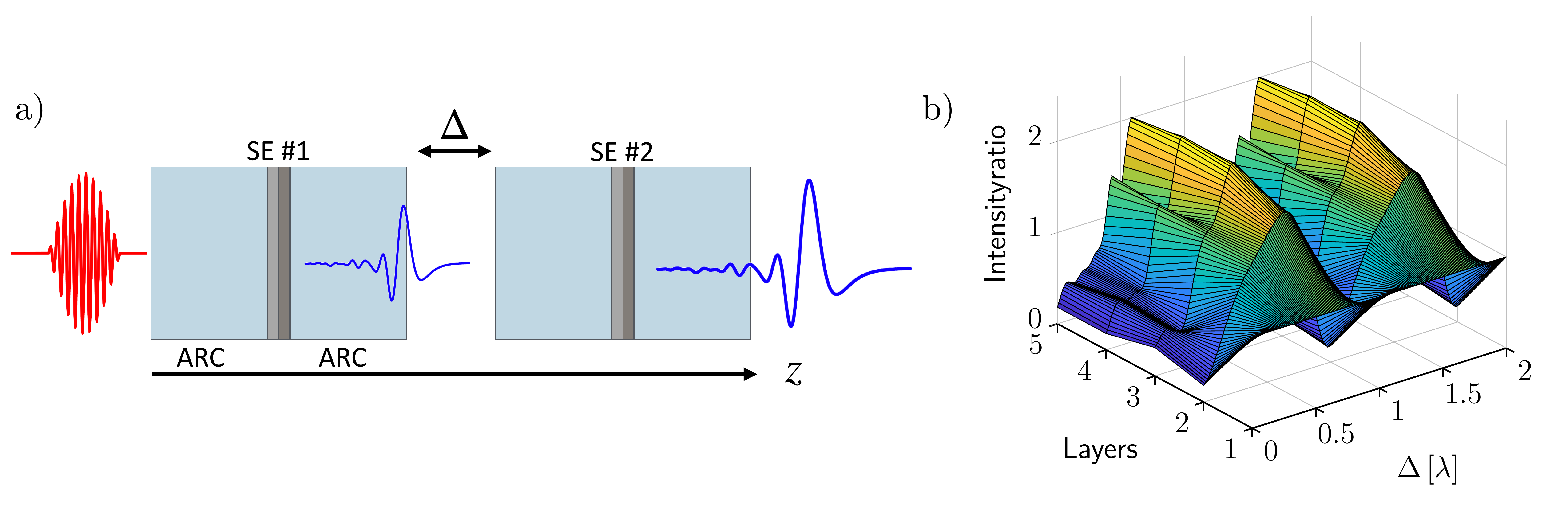}
	\caption{Integrated structure with anti-reflective coating (ARC) and multiple spintronic emitters (SE) in series (a). The distance between ARC-SE-ARC structures is denoted by $\Delta$. For varying distance and increasing number of layers, the intensity normalized to the emission of a single structure is plotted (b).\label{Fig3}}
\end{figure*}
%%%%%%%%%%%%%%%%%%%%%
% Results - General %
%%%%%%%%%%%%%%%%%%%%%
\section{Results}
\subsection{General Results}

We consider first a Fe(10\,nm)Pt(5\,nm) structure surrounded by vacuum and irradiated by an infrared laser pulse. We choose first a pulse length of 10\,fs, centered at 25\,fs, and a photon energy of 1.5\,eV~\cite{Seifert:2016kc}.  Subsequently, the pulse length will be varied to study its effect on the emitted spectrum. The laser excitation conditions are the same as in Ref.~\onlinecite{Nenno:KNmjgCOZ}.

Fig.~\ref{Fig1a}(b)-(c) show the calculated temporal profile of the spin current in platinum, as well as its spectrum. The spin current shows a characteristic bipolar shape, with a maximum close to the maximum of the laser pulse intensity at $25$\,fs. We explain the bipolar shape as follows~\cite{Nenno:KNmjgCOZ}. Carriers with spin polarization are excited in iron and start move into the platinum slab, inducing a positive spin current. However, after a very short time ($\approx 10$\,fs), carriers traverse the whole layer and change their propagation direction, effectively canceling any current. It is the small difference in propagating electron distribution undergoing scattering that generates an effective current. This current is dominated by majority carriers that enter the slab early on. When these majority carriers decay, the signal changes sign. The less intense tail of the spin current is then due to remaining minority carriers. The induced spin-current signal duration of approx. 150\,fs for a laser pulse of 10\,fs duration. For direct emission from the non-magnetic platinum layer, the spectrum is directly proportional to the charge current and therefore by Eq.~\eqref{eq:spinhall} the spin current.

%We want to note, that the obtained spectrum does not correspond to the measured spectrum in most experiments. Propagation through a dispersive
%medium as MgO and the detector response can alter the shape of the measured field considerably. We do not include these effects, since they strongly
%depend on the particular experimental set-up, however, we want to note that optical transfer-matrices are suitable to calculate propagation effects.

\subsection{Laser Pulse Duration}
In Fig.~\ref{Fig2}(a) shows emitted THz spectra for different temporal widths of the exciting laser pulse. We vary the FWHM of the pulse between 10\,fs and 100\,fs, covering the range around relevant scattering processes (e.g.~$\tau_{\mathrm{el}}=30$\,fs). We find two notable effects. The spectral maximum gets smaller and moves to lower frequencies for longer pulse duration. Fig.~\ref{Fig2}(b) shows both quantities plotted for all simulated excitation pulses. Longer excitation pulses induce hot-carrier distributions that persist over a longer period of time and thus, the resulting transport and relaxation dynamics result in a broader spin current with smaller magnitude (not shown). Hence, the spectrum of the emitted pulses peak at lower frequencies and with a smaller amplitude.

The second consequence of increasing pulse length is that the spectrum of the emitted THz radiation becomes narrower. In a simplified picture, the duration of the excitation would trigger dynamics on the same timescale and thus the spectral width of the emitted field should be proportional to the inverse of that duration. To analyze this in greater depth, we show the spectral width at 50\% of its maximum for all laser pulse durations in Fig.~\ref{Fig2}(c). The results are compared to the estimate $1/\mathrm{FWHM}$. We find that for short excitation lengths, the simple estimate deviates drastically from the simulated results, whereas they agree much better for longer laser pulses.

We note that the simulation of longer pulses needs to include thermalized spin currents which are not accessible in our current theoretical framework. The dynamics of shorter pulses can, in principle, be described, but such pulses typically have intensities that trigger demagnetization dynamics outside the scope of a linear charge accumulation.

%%%%%%%%%%%%%%%%%%%%%%%%
% Results - Multilayer %
%%%%%%%%%%%%%%%%%%%%%%%%
\subsection{Multilayer}

The energy of a single laser pulse is not completely absorbed within one emitter. For a thin emitter, e.g. Fe(3\,nm)Pt(2\,nm), approx.~40\% is absorbed for normal incidence from either side of the structure. Based on this observation, we propose an optimized structure of serial spintronic emitters (SE) coated with an anti-reflective coating (ARC) optimized to improve emission of a fixed THz frequency. The goal is to improve the power conversion rate and propose a structure suitable for on-chip terahertz opto-spintronic devices~\cite{2018arXiv180905391W}. Our model allows one to easily calculate the absorbed energy in these more complex structures, as well as the propagation and enhancement of the emitted radiation. We optimize the output for 4\,THz, close the maximum of the spectrum for an incident laser pulse of 70\,fs. The ARC is assumed to be non-absorbing with $n=3$
(polycrystalline diamond, for example, has $\tilde{n}_{THz} \approx 2.3 + i 3 \times 10^{-4}$, see Ref.~\onlinecite{dore-etal.98} or $n\approx 3.0-3.4$ for crystalline Sapphire and Silicon, see e.g.~Ref.~\onlinecite{Grischkowsky:90}).
 Its thickness is fixed to $\lambda/4$. In general, an optimization with respect to the specific material is possible. In this study, we fix the configuration of the spintronic emitter, the optical parameters of the ARC and vary the spacing between ARC--SE--ARC segments and their number. Fig.~\ref{Fig3} shows a sketch of the proposed heterostructure and the ratio between single emitter peak intensity and output of structures with a different number of spintronic emitter layers  for varying distance between the emitters.
 We see that a ratio of almost 2 can be obtained with 3 to 4 segments spaced at $\Delta \approx \lambda /2$. \\

%%%%%%%%%%%%%%
% Conclusion %
%%%%%%%%%%%%%%
\section{Conclusion}
We presented a multiscale theoretical framework for the processes that contribute to THz emission from spintronic heterostructures excited by optical pulses. We presented a microscopic analysis of optically induced spin- and charge currents that determine a spectrum in the THz range, which is directly proportional to the emitted pulse spectrum in the thin-film limit. Intrinsic material timescales in the range of THz oscillations are key for the physical processes that convert power from the exciting laser field to the THz emission. Our theory is based on a flexible solution of the BTE and does not contain any fit parameters and only a few well-established transport parameters from \textit{ab initio} calculations. The model can be extended to different structures with more challenging interface effects~\cite{PhysRevLett.120.207207,PhysRevApplied.10.044065} or ferrimagnetic emitter layers~\cite{doi:10.1021/acsphotonics.8b00839,Seifert_SPIN} by enhancing or replacing our method for calculating the electron dynamics in future studies. We showed that up to a number of four layers, a serial structure can result in an increased input- to output power ratio. The study can easily be extended further to perform an automated optimization with respect to the used material and laser parameters and sample configurations.

%%%%%%%%%%%%%%%%%%%
% Acknowledgments %
%%%%%%%%%%%%%%%%%%%
\begin{acknowledgments}
We are grateful for discussions with Garik Torosyan, Ren\'e Beigang, Evangelos Papaiouannou, Sascha Keller and Laura Scheuer. This work was supported by the German Science Foundation in the framework of the SFB/TRR 173 Spin+X (Project B03).
D.~M.~N.~thankfully acknowledges financial support from the Graduate School of Excellence MAINZ (Excellence Initiative DFG/GSC 266).
\end{acknowledgments}

%%%%%%%%%%%%%%
% References %
%%%%%%%%%%%%%%
\bibliography{References}

\begin{thebibliography}{24}%
\makeatletter
\providecommand \@ifxundefined [1]{%
 \@ifx{#1\undefined}
}%
\providecommand \@ifnum [1]{%
 \ifnum #1\expandafter \@firstoftwo
 \else \expandafter \@secondoftwo
 \fi
}%
\providecommand \@ifx [1]{%
 \ifx #1\expandafter \@firstoftwo
 \else \expandafter \@secondoftwo
 \fi
}%
\providecommand \natexlab [1]{#1}%
\providecommand \enquote  [1]{``#1''}%
\providecommand \bibnamefont  [1]{#1}%
\providecommand \bibfnamefont [1]{#1}%
\providecommand \citenamefont [1]{#1}%
\providecommand \href@noop [0]{\@secondoftwo}%
\providecommand \href [0]{\begingroup \@sanitize@url \@href}%
\providecommand \@href[1]{\@@startlink{#1}\@@href}%
\providecommand \@@href[1]{\endgroup#1\@@endlink}%
\providecommand \@sanitize@url [0]{\catcode `\\12\catcode `\$12\catcode
  `\&12\catcode `\#12\catcode `\^12\catcode `\_12\catcode `\%12\relax}%
\providecommand \@@startlink[1]{}%
\providecommand \@@endlink[0]{}%
\providecommand \url  [0]{\begingroup\@sanitize@url \@url }%
\providecommand \@url [1]{\endgroup\@href {#1}{\urlprefix }}%
\providecommand \urlprefix  [0]{URL }%
\providecommand \Eprint [0]{\href }%
\providecommand \doibase [0]{http://dx.doi.org/}%
\providecommand \selectlanguage [0]{\@gobble}%
\providecommand \bibinfo  [0]{\@secondoftwo}%
\providecommand \bibfield  [0]{\@secondoftwo}%
\providecommand \translation [1]{[#1]}%
\providecommand \BibitemOpen [0]{}%
\providecommand \bibitemStop [0]{}%
\providecommand \bibitemNoStop [0]{.\EOS\space}%
\providecommand \EOS [0]{\spacefactor3000\relax}%
\providecommand \BibitemShut  [1]{\csname bibitem#1\endcsname}%
\let\auto@bib@innerbib\@empty
%</preamble>
\bibitem [{\citenamefont {Kampfrath}\ \emph {et~al.}(2013)\citenamefont
  {Kampfrath}, \citenamefont {Battiato}, \citenamefont {Maldonado},
  \citenamefont {Eilers}, \citenamefont {N{\"o}tzold}, \citenamefont
  {M{\"a}hrlein}, \citenamefont {Zbarsky}, \citenamefont {Freimuth},
  \citenamefont {Mokrousov}, \citenamefont {Bl{\"u}gel}, \citenamefont {Wolf},
  \citenamefont {Radu}, \citenamefont {Oppeneer},\ and\ \citenamefont
  {M{\"u}nzenberg}}]{Kampfrath:2012kw}%
  \BibitemOpen
  \bibfield  {author} {\bibinfo {author} {\bibfnamefont {T.}~\bibnamefont
  {Kampfrath}}, \bibinfo {author} {\bibfnamefont {M.}~\bibnamefont {Battiato}},
  \bibinfo {author} {\bibfnamefont {P.}~\bibnamefont {Maldonado}}, \bibinfo
  {author} {\bibfnamefont {G.}~\bibnamefont {Eilers}}, \bibinfo {author}
  {\bibfnamefont {J.}~\bibnamefont {N{\"o}tzold}}, \bibinfo {author}
  {\bibfnamefont {S.}~\bibnamefont {M{\"a}hrlein}}, \bibinfo {author}
  {\bibfnamefont {V.}~\bibnamefont {Zbarsky}}, \bibinfo {author} {\bibfnamefont
  {F.}~\bibnamefont {Freimuth}}, \bibinfo {author} {\bibfnamefont
  {Y.}~\bibnamefont {Mokrousov}}, \bibinfo {author} {\bibfnamefont
  {S.}~\bibnamefont {Bl{\"u}gel}}, \bibinfo {author} {\bibfnamefont
  {M.}~\bibnamefont {Wolf}}, \bibinfo {author} {\bibfnamefont {I.}~\bibnamefont
  {Radu}}, \bibinfo {author} {\bibfnamefont {P.~M.}\ \bibnamefont {Oppeneer}},
  \ and\ \bibinfo {author} {\bibfnamefont {M.}~\bibnamefont {M{\"u}nzenberg}},\
  }\href@noop {} {\bibfield  {journal} {\bibinfo  {journal} {Nature
  Nanotechnology}\ }\textbf {\bibinfo {volume} {8}},\ \bibinfo {pages} {256}
  (\bibinfo {year} {2013})}\BibitemShut {NoStop}%
\bibitem [{\citenamefont {Seifert}\ \emph {et~al.}(2016)\citenamefont
  {Seifert}, \citenamefont {Jaiswal}, \citenamefont {Martens}, \citenamefont
  {Hannegan}, \citenamefont {Braun}, \citenamefont {Maldonado}, \citenamefont
  {Freimuth}, \citenamefont {Kronenberg}, \citenamefont {Henrizi},
  \citenamefont {Radu}, \citenamefont {Beaurepaire}, \citenamefont {Mokrousov},
  \citenamefont {Oppeneer}, \citenamefont {Jourdan}, \citenamefont {Jakob},
  \citenamefont {Turchinovich}, \citenamefont {Hayden}, \citenamefont {Wolf},
  \citenamefont {M{\"u}nzenberg}, \citenamefont {Kl{\"a}ui},\ and\
  \citenamefont {Kampfrath}}]{Seifert:2016kc}%
  \BibitemOpen
  \bibfield  {author} {\bibinfo {author} {\bibfnamefont {T.}~\bibnamefont
  {Seifert}}, \bibinfo {author} {\bibfnamefont {S.}~\bibnamefont {Jaiswal}},
  \bibinfo {author} {\bibfnamefont {U.}~\bibnamefont {Martens}}, \bibinfo
  {author} {\bibfnamefont {J.}~\bibnamefont {Hannegan}}, \bibinfo {author}
  {\bibfnamefont {L.}~\bibnamefont {Braun}}, \bibinfo {author} {\bibfnamefont
  {P.}~\bibnamefont {Maldonado}}, \bibinfo {author} {\bibfnamefont
  {F.}~\bibnamefont {Freimuth}}, \bibinfo {author} {\bibfnamefont
  {A.}~\bibnamefont {Kronenberg}}, \bibinfo {author} {\bibfnamefont
  {J.}~\bibnamefont {Henrizi}}, \bibinfo {author} {\bibfnamefont
  {I.}~\bibnamefont {Radu}}, \bibinfo {author} {\bibfnamefont {E.}~\bibnamefont
  {Beaurepaire}}, \bibinfo {author} {\bibfnamefont {Y.}~\bibnamefont
  {Mokrousov}}, \bibinfo {author} {\bibfnamefont {P.~M.}\ \bibnamefont
  {Oppeneer}}, \bibinfo {author} {\bibfnamefont {M.}~\bibnamefont {Jourdan}},
  \bibinfo {author} {\bibfnamefont {G.}~\bibnamefont {Jakob}}, \bibinfo
  {author} {\bibfnamefont {D.}~\bibnamefont {Turchinovich}}, \bibinfo {author}
  {\bibfnamefont {L.~M.}\ \bibnamefont {Hayden}}, \bibinfo {author}
  {\bibfnamefont {M.}~\bibnamefont {Wolf}}, \bibinfo {author} {\bibfnamefont
  {M.}~\bibnamefont {M{\"u}nzenberg}}, \bibinfo {author} {\bibfnamefont
  {M.}~\bibnamefont {Kl{\"a}ui}}, \ and\ \bibinfo {author} {\bibfnamefont
  {T.}~\bibnamefont {Kampfrath}},\ }\href@noop {} {\bibfield  {journal}
  {\bibinfo  {journal} {Nature Photonics}\ }\textbf {\bibinfo {volume} {10}},\
  \bibinfo {pages} {483} (\bibinfo {year} {2016})}\BibitemShut {NoStop}%
\bibitem [{\citenamefont {Cramer}\ \emph {et~al.}(2018)\citenamefont {Cramer},
  \citenamefont {Seifert}, \citenamefont {Kronenberg}, \citenamefont
  {Fuhrmann}, \citenamefont {Jakob}, \citenamefont {Jourdan}, \citenamefont
  {Kampfrath},\ and\ \citenamefont {Kl\"aui}}]{acs.nanolett.7b04538}%
  \BibitemOpen
  \bibfield  {author} {\bibinfo {author} {\bibfnamefont {J.}~\bibnamefont
  {Cramer}}, \bibinfo {author} {\bibfnamefont {T.}~\bibnamefont {Seifert}},
  \bibinfo {author} {\bibfnamefont {A.}~\bibnamefont {Kronenberg}}, \bibinfo
  {author} {\bibfnamefont {F.}~\bibnamefont {Fuhrmann}}, \bibinfo {author}
  {\bibfnamefont {G.}~\bibnamefont {Jakob}}, \bibinfo {author} {\bibfnamefont
  {M.}~\bibnamefont {Jourdan}}, \bibinfo {author} {\bibfnamefont
  {T.}~\bibnamefont {Kampfrath}}, \ and\ \bibinfo {author} {\bibfnamefont
  {M.}~\bibnamefont {Kl\"aui}},\ }\href@noop {} {\bibfield  {journal} {\bibinfo
   {journal} {Nano Letters}\ }\textbf {\bibinfo {volume} {18}},\ \bibinfo
  {pages} {1064} (\bibinfo {year} {2018})}\BibitemShut {NoStop}%
\bibitem [{\citenamefont {Torosyan}\ \emph {et~al.}(2018)\citenamefont
  {Torosyan}, \citenamefont {Keller}, \citenamefont {Scheuer}, \citenamefont
  {Beigang},\ and\ \citenamefont {Papaioannou}}]{Torosyan:2018iv}%
  \BibitemOpen
  \bibfield  {author} {\bibinfo {author} {\bibfnamefont {G.}~\bibnamefont
  {Torosyan}}, \bibinfo {author} {\bibfnamefont {S.}~\bibnamefont {Keller}},
  \bibinfo {author} {\bibfnamefont {L.}~\bibnamefont {Scheuer}}, \bibinfo
  {author} {\bibfnamefont {R.}~\bibnamefont {Beigang}}, \ and\ \bibinfo
  {author} {\bibfnamefont {E.~T.}\ \bibnamefont {Papaioannou}},\ }\href@noop {}
  {\bibfield  {journal} {\bibinfo  {journal} {Scientific Reports}\ }\textbf
  {\bibinfo {volume} {8}},\ \bibinfo {pages} {1311} (\bibinfo {year}
  {2018})}\BibitemShut {NoStop}%
\bibitem [{\citenamefont {{Papaioannou}}\ \emph {et~al.}(2018)\citenamefont
  {{Papaioannou}}, \citenamefont {{Torosyan}}, \citenamefont {{Keller}},
  \citenamefont {{Scheuer}}, \citenamefont {{Battiato}}, \citenamefont
  {{Katrine Mag-usara}}, \citenamefont {{L'huillier}}, \citenamefont {{Tani}},\
  and\ \citenamefont {{Beigang}}}]{2018arXiv180308838P}%
  \BibitemOpen
  \bibfield  {author} {\bibinfo {author} {\bibfnamefont {E.~T.}\ \bibnamefont
  {{Papaioannou}}}, \bibinfo {author} {\bibfnamefont {G.}~\bibnamefont
  {{Torosyan}}}, \bibinfo {author} {\bibfnamefont {S.}~\bibnamefont
  {{Keller}}}, \bibinfo {author} {\bibfnamefont {L.}~\bibnamefont {{Scheuer}}},
  \bibinfo {author} {\bibfnamefont {M.}~\bibnamefont {{Battiato}}}, \bibinfo
  {author} {\bibfnamefont {V.}~\bibnamefont {{Katrine Mag-usara}}}, \bibinfo
  {author} {\bibfnamefont {J.}~\bibnamefont {{L'huillier}}}, \bibinfo {author}
  {\bibfnamefont {M.}~\bibnamefont {{Tani}}}, \ and\ \bibinfo {author}
  {\bibfnamefont {R.}~\bibnamefont {{Beigang}}},\ }\href@noop {} {\bibfield
  {journal} {\bibinfo  {journal} {ArXiv e-prints}\ } (\bibinfo {year}
  {2018})},\ \Eprint {http://arxiv.org/abs/1803.08838} {arXiv:1803.08838}
  \BibitemShut {NoStop}%
\bibitem [{\citenamefont {{Herapath}}\ \emph {et~al.}(2018)\citenamefont
  {{Herapath}}, \citenamefont {{Hornett}}, \citenamefont {{Seifert}},
  \citenamefont {{Jakob}}, \citenamefont {{Kl{\"a}ui}}, \citenamefont
  {{Bertolotti}}, \citenamefont {{Kampfrath}},\ and\ \citenamefont
  {{Hendry}}}]{2018arXiv180800746H}%
  \BibitemOpen
  \bibfield  {author} {\bibinfo {author} {\bibfnamefont {R.~I.}\ \bibnamefont
  {{Herapath}}}, \bibinfo {author} {\bibfnamefont {S.~M.}\ \bibnamefont
  {{Hornett}}}, \bibinfo {author} {\bibfnamefont {T.}~\bibnamefont
  {{Seifert}}}, \bibinfo {author} {\bibfnamefont {G.}~\bibnamefont {{Jakob}}},
  \bibinfo {author} {\bibfnamefont {M.}~\bibnamefont {{Kl{\"a}ui}}}, \bibinfo
  {author} {\bibfnamefont {J.}~\bibnamefont {{Bertolotti}}}, \bibinfo {author}
  {\bibfnamefont {T.}~\bibnamefont {{Kampfrath}}}, \ and\ \bibinfo {author}
  {\bibfnamefont {E.}~\bibnamefont {{Hendry}}},\ }\href@noop {} {\bibfield
  {journal} {\bibinfo  {journal} {ArXiv e-prints}\ } (\bibinfo {year}
  {2018})},\ \Eprint {http://arxiv.org/abs/1808.00746} {arXiv:1808.00746
  [physics.optics]} \BibitemShut {NoStop}%
\bibitem [{\citenamefont {Nenno}\ \emph {et~al.}(2016)\citenamefont {Nenno},
  \citenamefont {Kaltenborn},\ and\ \citenamefont {Schneider}}]{Nenno:2016cm}%
  \BibitemOpen
  \bibfield  {author} {\bibinfo {author} {\bibfnamefont {D.~M.}\ \bibnamefont
  {Nenno}}, \bibinfo {author} {\bibfnamefont {S.}~\bibnamefont {Kaltenborn}}, \
  and\ \bibinfo {author} {\bibfnamefont {H.~C.}\ \bibnamefont {Schneider}},\
  }\href@noop {} {\bibfield  {journal} {\bibinfo  {journal} {Physical Review
  B}\ }\textbf {\bibinfo {volume} {94}},\ \bibinfo {pages} {115102} (\bibinfo
  {year} {2016})}\BibitemShut {NoStop}%
\bibitem [{\citenamefont {Hurst}\ \emph {et~al.}(2018)\citenamefont {Hurst},
  \citenamefont {Hervieux},\ and\ \citenamefont {Manfredi}}]{Hurst:2018bz}%
  \BibitemOpen
  \bibfield  {author} {\bibinfo {author} {\bibfnamefont {J.}~\bibnamefont
  {Hurst}}, \bibinfo {author} {\bibfnamefont {P.-A.}\ \bibnamefont {Hervieux}},
  \ and\ \bibinfo {author} {\bibfnamefont {G.}~\bibnamefont {Manfredi}},\
  }\href@noop {} {\bibfield  {journal} {\bibinfo  {journal} {Physical Review
  B}\ }\textbf {\bibinfo {volume} {97}},\ \bibinfo {pages} {014424} (\bibinfo
  {year} {2018})}\BibitemShut {NoStop}%
\bibitem [{\citenamefont {Manfredi}\ and\ \citenamefont
  {Hervieux}(2005)}]{Manfredi:2005ba}%
  \BibitemOpen
  \bibfield  {author} {\bibinfo {author} {\bibfnamefont {G.}~\bibnamefont
  {Manfredi}}\ and\ \bibinfo {author} {\bibfnamefont {P.~A.}\ \bibnamefont
  {Hervieux}},\ }\href@noop {} {\bibfield  {journal} {\bibinfo  {journal}
  {Physical Review B}\ }\textbf {\bibinfo {volume} {72}},\ \bibinfo {pages}
  {155421} (\bibinfo {year} {2005})}\BibitemShut {NoStop}%
\bibitem [{\citenamefont {Nenno}\ \emph {et~al.}(2018)\citenamefont {Nenno},
  \citenamefont {Rethfeld},\ and\ \citenamefont {Schneider}}]{Nenno:KNmjgCOZ}%
  \BibitemOpen
  \bibfield  {author} {\bibinfo {author} {\bibfnamefont {D.~M.}\ \bibnamefont
  {Nenno}}, \bibinfo {author} {\bibfnamefont {B.}~\bibnamefont {Rethfeld}}, \
  and\ \bibinfo {author} {\bibfnamefont {H.~C.}\ \bibnamefont {Schneider}},\
  }\href {\doibase 10.1103/PhysRevB.98.224416} {\bibfield  {journal} {\bibinfo
  {journal} {Phys. Rev. B}\ }\textbf {\bibinfo {volume} {98}},\ \bibinfo
  {pages} {224416} (\bibinfo {year} {2018})}\BibitemShut {NoStop}%
\bibitem [{\citenamefont {Prentice}(2000)}]{Prentince_JOPD_2000}%
  \BibitemOpen
  \bibfield  {author} {\bibinfo {author} {\bibfnamefont {J.~S.~C.}\
  \bibnamefont {Prentice}},\ }\href
  {http://stacks.iop.org/0022-3727/33/i=24/a=302} {\bibfield  {journal}
  {\bibinfo  {journal} {Journal of Physics D: Applied Physics}\ }\textbf
  {\bibinfo {volume} {33}},\ \bibinfo {pages} {3139} (\bibinfo {year}
  {2000})}\BibitemShut {NoStop}%
\bibitem [{\citenamefont {Werner}\ \emph {et~al.}(2009)\citenamefont {Werner},
  \citenamefont {Glantschnig},\ and\ \citenamefont
  {Ambrosch-Draxl}}]{doi:10.1063/1.3243762}%
  \BibitemOpen
  \bibfield  {author} {\bibinfo {author} {\bibfnamefont {W.~S.~M.}\
  \bibnamefont {Werner}}, \bibinfo {author} {\bibfnamefont {K.}~\bibnamefont
  {Glantschnig}}, \ and\ \bibinfo {author} {\bibfnamefont {C.}~\bibnamefont
  {Ambrosch-Draxl}},\ }\href {\doibase 10.1063/1.3243762} {\bibfield  {journal}
  {\bibinfo  {journal} {Journal of Physical and Chemical Reference Data}\
  }\textbf {\bibinfo {volume} {38}},\ \bibinfo {pages} {1013} (\bibinfo {year}
  {2009})}\BibitemShut {NoStop}%
\bibitem [{\citenamefont {Koopmans}\ \emph {et~al.}(2000)\citenamefont
  {Koopmans}, \citenamefont {van Kampen}, \citenamefont {Kohlhepp},\ and\
  \citenamefont {de~Jonge}}]{PhysRevLett.85.844}%
  \BibitemOpen
  \bibfield  {author} {\bibinfo {author} {\bibfnamefont {B.}~\bibnamefont
  {Koopmans}}, \bibinfo {author} {\bibfnamefont {M.}~\bibnamefont {van
  Kampen}}, \bibinfo {author} {\bibfnamefont {J.~T.}\ \bibnamefont {Kohlhepp}},
  \ and\ \bibinfo {author} {\bibfnamefont {W.~J.~M.}\ \bibnamefont
  {de~Jonge}},\ }\href {\doibase 10.1103/PhysRevLett.85.844} {\bibfield
  {journal} {\bibinfo  {journal} {Phys. Rev. Lett.}\ }\textbf {\bibinfo
  {volume} {85}},\ \bibinfo {pages} {844} (\bibinfo {year} {2000})}\BibitemShut
  {NoStop}%
\bibitem [{\citenamefont {Battiato}\ \emph {et~al.}(2010)\citenamefont
  {Battiato}, \citenamefont {Carva},\ and\ \citenamefont
  {Oppeneer}}]{Battiato:2010br}%
  \BibitemOpen
  \bibfield  {author} {\bibinfo {author} {\bibfnamefont {M.}~\bibnamefont
  {Battiato}}, \bibinfo {author} {\bibfnamefont {K.}~\bibnamefont {Carva}}, \
  and\ \bibinfo {author} {\bibfnamefont {P.~M.}\ \bibnamefont {Oppeneer}},\
  }\href@noop {} {\bibfield  {journal} {\bibinfo  {journal} {Physical Review
  Letters}\ }\textbf {\bibinfo {volume} {105}},\ \bibinfo {pages} {027203}
  (\bibinfo {year} {2010})}\BibitemShut {NoStop}%
\bibitem [{\citenamefont {Zhukov}\ \emph {et~al.}(2006)\citenamefont {Zhukov},
  \citenamefont {Chulkov},\ and\ \citenamefont {Echenique}}]{Zhukov:2006ky}%
  \BibitemOpen
  \bibfield  {author} {\bibinfo {author} {\bibfnamefont {V.~P.}\ \bibnamefont
  {Zhukov}}, \bibinfo {author} {\bibfnamefont {E.~V.}\ \bibnamefont {Chulkov}},
  \ and\ \bibinfo {author} {\bibfnamefont {P.~M.}\ \bibnamefont {Echenique}},\
  }\href@noop {} {\bibfield  {journal} {\bibinfo  {journal} {Physical Review
  B}\ }\textbf {\bibinfo {volume} {73}},\ \bibinfo {pages} {465} (\bibinfo
  {year} {2006})}\BibitemShut {NoStop}%
\bibitem [{\citenamefont {Kaltenborn}\ and\ \citenamefont
  {Schneider}(2014)}]{Kaltenborn:2014du}%
  \BibitemOpen
  \bibfield  {author} {\bibinfo {author} {\bibfnamefont {S.}~\bibnamefont
  {Kaltenborn}}\ and\ \bibinfo {author} {\bibfnamefont {H.~C.}\ \bibnamefont
  {Schneider}},\ }\href@noop {} {\bibfield  {journal} {\bibinfo  {journal}
  {Physical Review B}\ }\textbf {\bibinfo {volume} {90}},\ \bibinfo {pages}
  {201104} (\bibinfo {year} {2014})}\BibitemShut {NoStop}%
\bibitem [{\citenamefont {Meystre}\ and\ \citenamefont
  {Sargent}(1991)}]{meystre-sargent.91}%
  \BibitemOpen
  \bibfield  {author} {\bibinfo {author} {\bibfnamefont {P.}~\bibnamefont
  {Meystre}}\ and\ \bibinfo {author} {\bibfnamefont {M.}~\bibnamefont
  {Sargent}},\ }\href@noop {} {\emph {\bibinfo {title} {Elements of Quantum
  Optics}}},\ \bibinfo {edition} {2nd}\ ed.\ (\bibinfo  {publisher}
  {Springer},\ \bibinfo {address} {New York},\ \bibinfo {year}
  {1991})\BibitemShut {NoStop}%
\bibitem [{\citenamefont {{Wu}}\ \emph {et~al.}(2018)\citenamefont {{Wu}},
  \citenamefont {{Nie}}, \citenamefont {{Wang}}, \citenamefont {{Xiao}},
  \citenamefont {{Kong}}, \citenamefont {{Pandey}}, \citenamefont {{Gao}},
  \citenamefont {{Wen}}, \citenamefont {{Zhao}}, \citenamefont {{Ruan}},
  \citenamefont {{Miao}}, \citenamefont {{Wang}}, \citenamefont {{Li}},\ and\
  \citenamefont {{Wang}}}]{2018arXiv180905391W}%
  \BibitemOpen
  \bibfield  {author} {\bibinfo {author} {\bibfnamefont {X.}~\bibnamefont
  {{Wu}}}, \bibinfo {author} {\bibfnamefont {T.}~\bibnamefont {{Nie}}},
  \bibinfo {author} {\bibfnamefont {B.}~\bibnamefont {{Wang}}}, \bibinfo
  {author} {\bibfnamefont {M.}~\bibnamefont {{Xiao}}}, \bibinfo {author}
  {\bibfnamefont {D.}~\bibnamefont {{Kong}}}, \bibinfo {author} {\bibfnamefont
  {C.}~\bibnamefont {{Pandey}}}, \bibinfo {author} {\bibfnamefont
  {Y.}~\bibnamefont {{Gao}}}, \bibinfo {author} {\bibfnamefont
  {L.}~\bibnamefont {{Wen}}}, \bibinfo {author} {\bibfnamefont
  {W.}~\bibnamefont {{Zhao}}}, \bibinfo {author} {\bibfnamefont
  {C.}~\bibnamefont {{Ruan}}}, \bibinfo {author} {\bibfnamefont
  {J.}~\bibnamefont {{Miao}}}, \bibinfo {author} {\bibfnamefont
  {L.}~\bibnamefont {{Wang}}}, \bibinfo {author} {\bibfnamefont
  {Y.}~\bibnamefont {{Li}}}, \ and\ \bibinfo {author} {\bibfnamefont {K.~L.}\
  \bibnamefont {{Wang}}},\ }\href@noop {} {\bibfield  {journal} {\bibinfo
  {journal} {ArXiv e-prints}\ } (\bibinfo {year} {2018})},\ \Eprint
  {http://arxiv.org/abs/1809.05391} {arXiv:1809.05391 [physics.optics]}
  \BibitemShut {NoStop}%
\bibitem [{\citenamefont {Dore}\ \emph {et~al.}(1998)\citenamefont {Dore},
  \citenamefont {Nucara}, \citenamefont {Cannavo}, \citenamefont {DeMarzi},
  \citenamefont {Calvani}, \citenamefont {Marcelli}, \citenamefont {Sussman},
  \citenamefont {Whitehead}, \citenamefont {Dodge}, \citenamefont {Krehan},\
  and\ \citenamefont {Peters}}]{dore-etal.98}%
  \BibitemOpen
  \bibfield  {author} {\bibinfo {author} {\bibfnamefont {P.}~\bibnamefont
  {Dore}}, \bibinfo {author} {\bibfnamefont {A.}~\bibnamefont {Nucara}},
  \bibinfo {author} {\bibfnamefont {D.}~\bibnamefont {Cannavo}}, \bibinfo
  {author} {\bibfnamefont {G.}~\bibnamefont {DeMarzi}}, \bibinfo {author}
  {\bibfnamefont {P.}~\bibnamefont {Calvani}}, \bibinfo {author} {\bibfnamefont
  {A.}~\bibnamefont {Marcelli}}, \bibinfo {author} {\bibfnamefont
  {R.}~\bibnamefont {Sussman}}, \bibinfo {author} {\bibfnamefont
  {A.}~\bibnamefont {Whitehead}}, \bibinfo {author} {\bibfnamefont
  {C.}~\bibnamefont {Dodge}}, \bibinfo {author} {\bibfnamefont
  {A.}~\bibnamefont {Krehan}}, \ and\ \bibinfo {author} {\bibfnamefont
  {H.}~\bibnamefont {Peters}},\ }\href@noop {} {\bibfield  {journal} {\bibinfo
  {journal} {Applied Optics}\ }\textbf {\bibinfo {volume} {37}},\ \bibinfo
  {pages} {5731 } (\bibinfo {year} {1998})}\BibitemShut {NoStop}%
\bibitem [{\citenamefont {Grischkowsky}\ \emph {et~al.}(1990)\citenamefont
  {Grischkowsky}, \citenamefont {Keiding}, \citenamefont {van Exter},\ and\
  \citenamefont {Fattinger}}]{Grischkowsky:90}%
  \BibitemOpen
  \bibfield  {author} {\bibinfo {author} {\bibfnamefont {D.}~\bibnamefont
  {Grischkowsky}}, \bibinfo {author} {\bibfnamefont {S.}~\bibnamefont
  {Keiding}}, \bibinfo {author} {\bibfnamefont {M.}~\bibnamefont {van Exter}},
  \ and\ \bibinfo {author} {\bibfnamefont {C.}~\bibnamefont {Fattinger}},\
  }\href {\doibase 10.1364/JOSAB.7.002006} {\bibfield  {journal} {\bibinfo
  {journal} {J. Opt. Soc. Am. B}\ }\textbf {\bibinfo {volume} {7}},\ \bibinfo
  {pages} {2006} (\bibinfo {year} {1990})}\BibitemShut {NoStop}%
\bibitem [{\citenamefont {Jungfleisch}\ \emph {et~al.}(2018)\citenamefont
  {Jungfleisch}, \citenamefont {Zhang}, \citenamefont {Zhang}, \citenamefont
  {Pearson}, \citenamefont {Schaller}, \citenamefont {Wen},\ and\ \citenamefont
  {Hoffmann}}]{PhysRevLett.120.207207}%
  \BibitemOpen
  \bibfield  {author} {\bibinfo {author} {\bibfnamefont {M.~B.}\ \bibnamefont
  {Jungfleisch}}, \bibinfo {author} {\bibfnamefont {Q.}~\bibnamefont {Zhang}},
  \bibinfo {author} {\bibfnamefont {W.}~\bibnamefont {Zhang}}, \bibinfo
  {author} {\bibfnamefont {J.~E.}\ \bibnamefont {Pearson}}, \bibinfo {author}
  {\bibfnamefont {R.~D.}\ \bibnamefont {Schaller}}, \bibinfo {author}
  {\bibfnamefont {H.}~\bibnamefont {Wen}}, \ and\ \bibinfo {author}
  {\bibfnamefont {A.}~\bibnamefont {Hoffmann}},\ }\href {\doibase
  10.1103/PhysRevLett.120.207207} {\bibfield  {journal} {\bibinfo  {journal}
  {Phys. Rev. Lett.}\ }\textbf {\bibinfo {volume} {120}},\ \bibinfo {pages}
  {207207} (\bibinfo {year} {2018})}\BibitemShut {NoStop}%
\bibitem [{\citenamefont {Dewhurst}\ \emph {et~al.}(2018)\citenamefont
  {Dewhurst}, \citenamefont {Shallcross}, \citenamefont {Gross},\ and\
  \citenamefont {Sharma}}]{PhysRevApplied.10.044065}%
  \BibitemOpen
  \bibfield  {author} {\bibinfo {author} {\bibfnamefont {J.~K.}\ \bibnamefont
  {Dewhurst}}, \bibinfo {author} {\bibfnamefont {S.}~\bibnamefont
  {Shallcross}}, \bibinfo {author} {\bibfnamefont {E.~K.~U.}\ \bibnamefont
  {Gross}}, \ and\ \bibinfo {author} {\bibfnamefont {S.}~\bibnamefont
  {Sharma}},\ }\href {\doibase 10.1103/PhysRevApplied.10.044065} {\bibfield
  {journal} {\bibinfo  {journal} {Phys. Rev. Applied}\ }\textbf {\bibinfo
  {volume} {10}},\ \bibinfo {pages} {044065} (\bibinfo {year}
  {2018})}\BibitemShut {NoStop}%
\bibitem [{\citenamefont {Schneider}\ \emph {et~al.}(2018)\citenamefont
  {Schneider}, \citenamefont {Fix}, \citenamefont {Heming}, \citenamefont
  {Michaelis~de Vasconcellos}, \citenamefont {Albrecht},\ and\ \citenamefont
  {Bratschitsch}}]{doi:10.1021/acsphotonics.8b00839}%
  \BibitemOpen
  \bibfield  {author} {\bibinfo {author} {\bibfnamefont {R.}~\bibnamefont
  {Schneider}}, \bibinfo {author} {\bibfnamefont {M.}~\bibnamefont {Fix}},
  \bibinfo {author} {\bibfnamefont {R.}~\bibnamefont {Heming}}, \bibinfo
  {author} {\bibfnamefont {S.}~\bibnamefont {Michaelis~de Vasconcellos}},
  \bibinfo {author} {\bibfnamefont {M.}~\bibnamefont {Albrecht}}, \ and\
  \bibinfo {author} {\bibfnamefont {R.}~\bibnamefont {Bratschitsch}},\
  }\href@noop {} {\bibfield  {journal} {\bibinfo  {journal} {ACS Photonics}\
  }\textbf {\bibinfo {volume} {5}},\ \bibinfo {pages} {3936} (\bibinfo {year}
  {2018})}\BibitemShut {NoStop}%
\bibitem [{\citenamefont {Seifert}\ \emph {et~al.}(2017)\citenamefont
  {Seifert}, \citenamefont {Martens}, \citenamefont {G\"unther}, \citenamefont
  {Schoen}, \citenamefont {Radu}, \citenamefont {Chen}, \citenamefont {Lucas},
  \citenamefont {Ramos}, \citenamefont {Aguirre}, \citenamefont {Algarabel},
  \citenamefont {Anadón}, \citenamefont {Körner}, \citenamefont {Walowski},
  \citenamefont {Back}, \citenamefont {Ibarra}, \citenamefont {Morellón},
  \citenamefont {Saitoh}, \citenamefont {Wolf}, \citenamefont {Song},
  \citenamefont {Uchida}, \citenamefont {Münzenberg}, \citenamefont {Radu},\
  and\ \citenamefont {Kampfrath}}]{Seifert_SPIN}%
  \BibitemOpen
  \bibfield  {author} {\bibinfo {author} {\bibfnamefont {T.}~\bibnamefont
  {Seifert}}, \bibinfo {author} {\bibfnamefont {U.}~\bibnamefont {Martens}},
  \bibinfo {author} {\bibfnamefont {S.}~\bibnamefont {G\"unther}}, \bibinfo
  {author} {\bibfnamefont {M.~A.~W.}\ \bibnamefont {Schoen}}, \bibinfo {author}
  {\bibfnamefont {F.}~\bibnamefont {Radu}}, \bibinfo {author} {\bibfnamefont
  {X.~Z.}\ \bibnamefont {Chen}}, \bibinfo {author} {\bibfnamefont
  {I.}~\bibnamefont {Lucas}}, \bibinfo {author} {\bibfnamefont
  {R.}~\bibnamefont {Ramos}}, \bibinfo {author} {\bibfnamefont {M.~H.}\
  \bibnamefont {Aguirre}}, \bibinfo {author} {\bibfnamefont {P.~A.}\
  \bibnamefont {Algarabel}}, \bibinfo {author} {\bibfnamefont {A.}~\bibnamefont
  {Anadón}}, \bibinfo {author} {\bibfnamefont {H.~S.}\ \bibnamefont
  {Körner}}, \bibinfo {author} {\bibfnamefont {J.}~\bibnamefont {Walowski}},
  \bibinfo {author} {\bibfnamefont {C.}~\bibnamefont {Back}}, \bibinfo {author}
  {\bibfnamefont {M.~R.}\ \bibnamefont {Ibarra}}, \bibinfo {author}
  {\bibfnamefont {L.}~\bibnamefont {Morellón}}, \bibinfo {author}
  {\bibfnamefont {E.}~\bibnamefont {Saitoh}}, \bibinfo {author} {\bibfnamefont
  {M.}~\bibnamefont {Wolf}}, \bibinfo {author} {\bibfnamefont {C.}~\bibnamefont
  {Song}}, \bibinfo {author} {\bibfnamefont {K.}~\bibnamefont {Uchida}},
  \bibinfo {author} {\bibfnamefont {M.}~\bibnamefont {Münzenberg}}, \bibinfo
  {author} {\bibfnamefont {I.}~\bibnamefont {Radu}}, \ and\ \bibinfo {author}
  {\bibfnamefont {T.}~\bibnamefont {Kampfrath}},\ }\href@noop {} {\bibfield
  {journal} {\bibinfo  {journal} {SPIN}\ }\textbf {\bibinfo {volume} {07}},\
  \bibinfo {pages} {1740010} (\bibinfo {year} {2017})}\BibitemShut {NoStop}%
\end{thebibliography}%
\end{document}